\documentclass[printer]{aa} 
\usepackage{graphicx}
\begin{document}
\title{Periodic H$\alpha$ variations in GL 581:\\ Further evidence for
an activity origin to GL 581d }


   \author{A. P. Hatzes
         }

   \offprints{
    Artie Hatzes, \email{artie@tls-tautenburg.de}\\$*$~
  \\}

     \institute{Th\"uringer Landessternwarte Tautenburg,
                Sternwarte 5, D-07778 Tautenburg, Germany
}

   \date{Received; accepted}

 
  \abstract
  {Radial velocity measurements initially showed evidence that the 
M dwarf GL 581 might 
    host up to six planetary companions. Two of these, GL 581g and GL 581d
   had orbital distances in the so-called ``habitable zone'' of the star. The existence of both of these
planets have been called into question. Additional radial velocity measurements
for GL 581g could not confirm its presence. A study of H$\alpha$ in GL 581 showed that changes in this
activity indicator 
   correlated with radial velocity variations attributed to
GL 581d.  Thus two planets that were important for studies of habitable planets may be artifacts
   of stellar activity. }  
   {Previous investigations analyzing the same activity
   data have reached different conclusions regarding the existence of GL 581d.
   We therefore investigated the  H$\alpha$ variations for GL 581 to assess the nature of the radial velocity 
   variations attributed to the possible planet GL 581d. }
   {We performed a Fourier analysis of the published H$\alpha$ measurements for GL 581d. Fourier components are selectively found
   and removed in a so-called pre-whitening process thus isolating any  variations at the orbital frequency of GL 581d. 
   }
   {The frequency analysis yields five significant frequencies, one of which is associated with the 66.7 d orbital period of the  presumed planet Gl 581d.
   The H$\alpha$ variations  at this period show sine-like variations that are 180$^\circ$ out-of-phase with the radial velocity variations of
   GL 581d. This is seen in the full data set that spans almost 7 years, as well as a subset of the data near the end
   of the time series that had  good temporal sampling over 230 days.
   No significant temporal variations are found in the ratio of the amplitudes of the H$\alpha$ index and radial velocity variations.
This provides additional
   evidence that the radial velocity signal attributed to GL 581d is in fact due to stellar activity. }
   {The analysis confirms the anti-correlation of the radial velocity of GL 581d with the H$\alpha$ equivalent width and provides additional  strong evidence that the signal of GL 581d is intrinsic to the star. 
}

\titlerunning{H$\alpha$ Variations in GL 581}
\maketitle

%

\section{Introduction}
Radial velocity (RV) measurements have demonstrated great  successful 
at discovering extrasolar planets. New instrumentation, 
improved calibration methods, and innovative analysis techniques have steadily improved
the RV precision to the point that we can routinely make measurements
with precisions of $\sim$ 1 m\,s$^{-1}$ or better. At this level of precision
the stellar intrinsic noise now represents a significant
contribution to the measurement ``error'', often referred to as the stellar RV
``jitter''. In the best case this  RV jitter can hinder the detection of planetary companions,
in the worse case it can create a periodic signal that is interpreted as arising from
a planetary companion.

M dwarf stars are  objects that 
have become particularly attractive for Doppler surveys because the RV amplitude
of the host star caused by an orbiting terrestrial planet in the habitable zone 
is $\sim$ few m\,s$^{-1}$, a value
easily measured by current techniques. Unfortunately, M dwarfs can be active 
and  the orbital period of a planet in the habitable zone
is days to weeks and this is comparable to the 
time scales of stellar activity (rotational modulation, spot evolution, etc.). Activity-related RV jitter
 may produce false planets. 

A case in point is the planetary system around the M dwarf star GL 581. {Mayor et al.
2009) reported four planet in the system and shortly afterwards Vogt et al. (2010) claimed 
six planet candidates  with periods { up to 433} d}.
Two of these planets,
GL 581d and GL 581g, were of particular interest because they had orbital distances
that placed them within the habitable zone of the star. Subsequent RV measurements
(Forveille et al. 2009) could not confirm the presence of GL 581g, although this
has been the subject of debate (Vogt et al. 2012). Hatzes (2014) demonstrated that the signal attributed
to  GL 581g was real and significant, but most likely due to activity as it was not coherent
on long time scales.

Baluev (2012) was the first to question the reality of GL 581d based on an analysis of the red noise in the RV data.
This doubt was
considerably strengthened through a 
study by  Robertson et al. (2014). They argued that
the 66-d orbital period of GL 581d was actually a harmonic of the 130 d rotation period of the
star. This was based on an apparent correlation between the RV variations due to
GL 581d and the equivalent width of H$\alpha$, their so-called $I_{H\alpha}$ index.
Correcting the RVs for this correlation
eliminated the signals attributed to GL 581d and GL 581g, while boosting the significance
of the other three planets. 

The nature of the 66-d RV period of GL 581 (hereafter referred to as the ``orbital period'' regardless of whether
the planet exists or not) still remains the subject of debate. Anglada-Escud{\'e}
\& Tuomi (2015) questioned the conclusions of Robertson et al. (2014), arguing that their result
came from an improper use of periodograms on the residual data. Although Anglada-Escud{\'e}
\& Tuomi agreed that there was a substantial correlation between the RV variations and $I_{H\alpha}$ index,
they argued that there was no clear evidence of time variability of this index.

It is important to establish the true nature of the RV variations of the purported planet GL 581d as this
has important consequences for RV searches for planets in the habitable zone of M dwarf stars.
This is especially true since different analyses of the same $I_{H\alpha}$ index measurements
 arrived at different conclusions.
Therefore, we investigated the H$\alpha$ variations using a different approach to the one taken by
Robertson et al. (2014), namely a  Fourier analysis of the frequency components in the
H$\alpha$ time series. The aim of this work is to confirm or refute the anti-correlation between RV and 
H$\alpha$ equivalent width found by Robertson et al. (2014).

\section{The Temporal Variations of H$\alpha$}

\subsection{Frequency Analysis}

\begin{figure}[h]
\resizebox{\hsize}{!}{\includegraphics{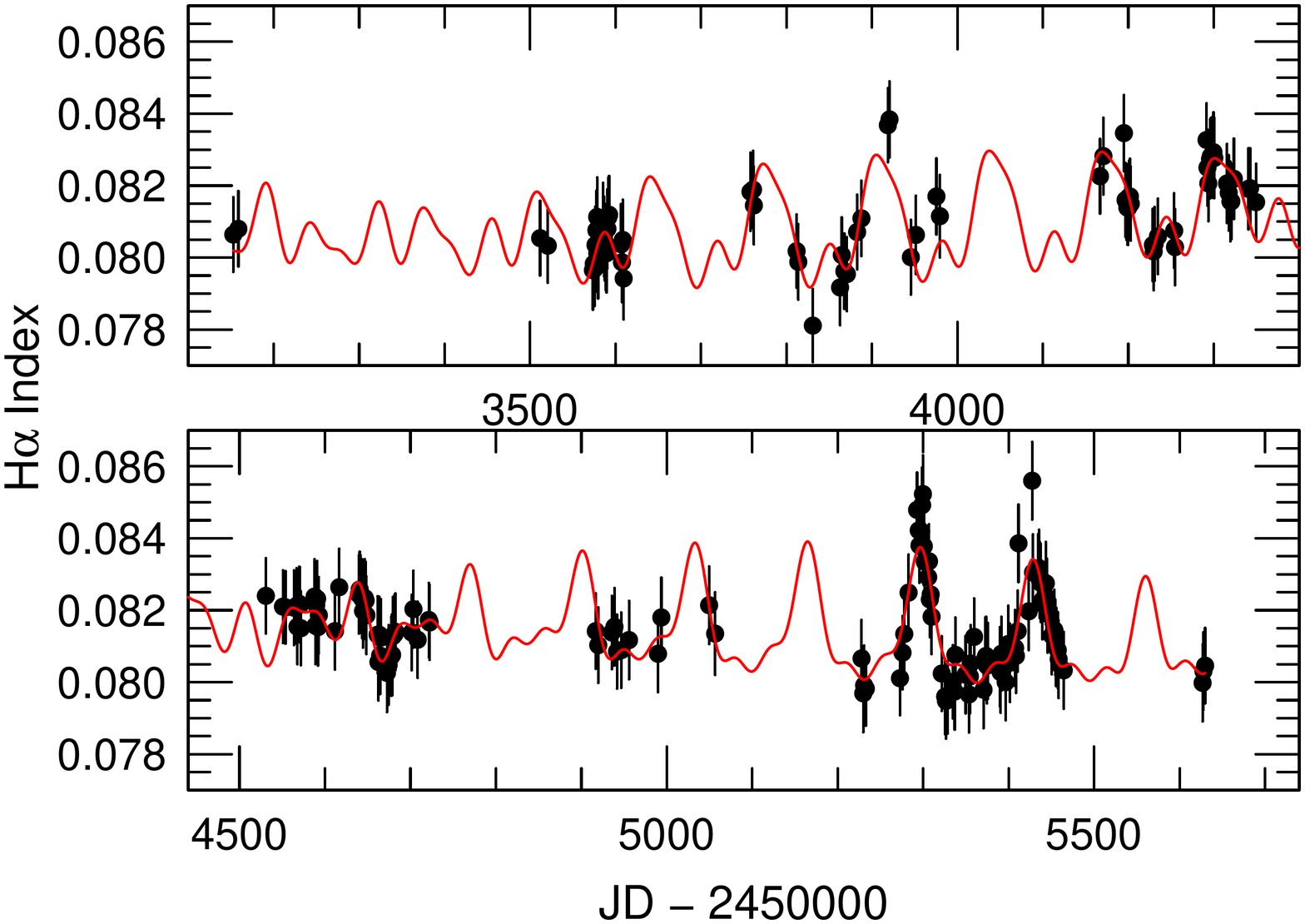}}
\caption{Time series of the H$\alpha$  equivalent width measurements for GL~581. The line represents
the five frequency fit using the entries in Table 1 ($f_1$ $-$ $f_5$).
}
\label{time}
\end{figure}

For this study we used the $I_{H\alpha}$ index  measurements of GL~581 from Robertson et al. (2014)
which are shown in Figure~\ref{time}. The data were analyzed using
a Fourier approach. We first found the dominant
component in the Fourier transform, and this was fit  
with a sine function of the appropriate period, amplitude,
and phase. The contribution of this frequency was then removed from the time
series  and we proceeded to the next dominant sine term in the residuals. This process
is often referred to as ``pre-whitening''.
The advantage of this technique is that since you are fitting and removing  a sine function
sampled in the same way as the data, alias frequencies are also removed.

Normally one associates pre-whitening with finding strictly periodic signals in your data, often in cases
associated with stellar oscillations. However, pre-whitening can be more versatile in that it finds the dominant
Fourier components that describe the overall variations in your time series, even if they do not appear to be periodic.
The mathematical foundation for this is that trigonometric functions form  basis set and a linear combination
of these provide an alternative mathematical representation of the series.  
Pre-whitening finds those Fourier components in the time series
that are clearly above the noise level. The sum of these may provide an adequate representation 
of the time variations in the RV due to activity.

For our pre-whitening analysis we used the program {\it Period04}  (Lenz \& Breger 2005). 
Table 1 lists the dominant frequencies (labeled $f_1$ -- $f_5$), equivalent periods, and amplitudes found by the Fourier analysis.
The pre-whitening procedure was stopped when the  dominant peak in the residuals
had a peak  less than four times the surrounding noise level. Peaks of this amplitude generally have a false alarm
probability of $\sim$1\% Kuschnig et al. (1997). 
Once the dominant frequency components were found
a simultaneous fit was made to the data optimizing the amplitude, period, and phase
of the individual sine terms. 
The multi-sine fit using these frequencies is shown as the curve in Figure~\ref{time}. 
The first two frequencies, $f_1$ and $f_2$ are associated with the $\sim$ 130 d rotation period of GL~581. The third frequency is the
orbital frequency of GL~518d.

\begin{figure}[h]
\resizebox{\hsize}{!}{\includegraphics{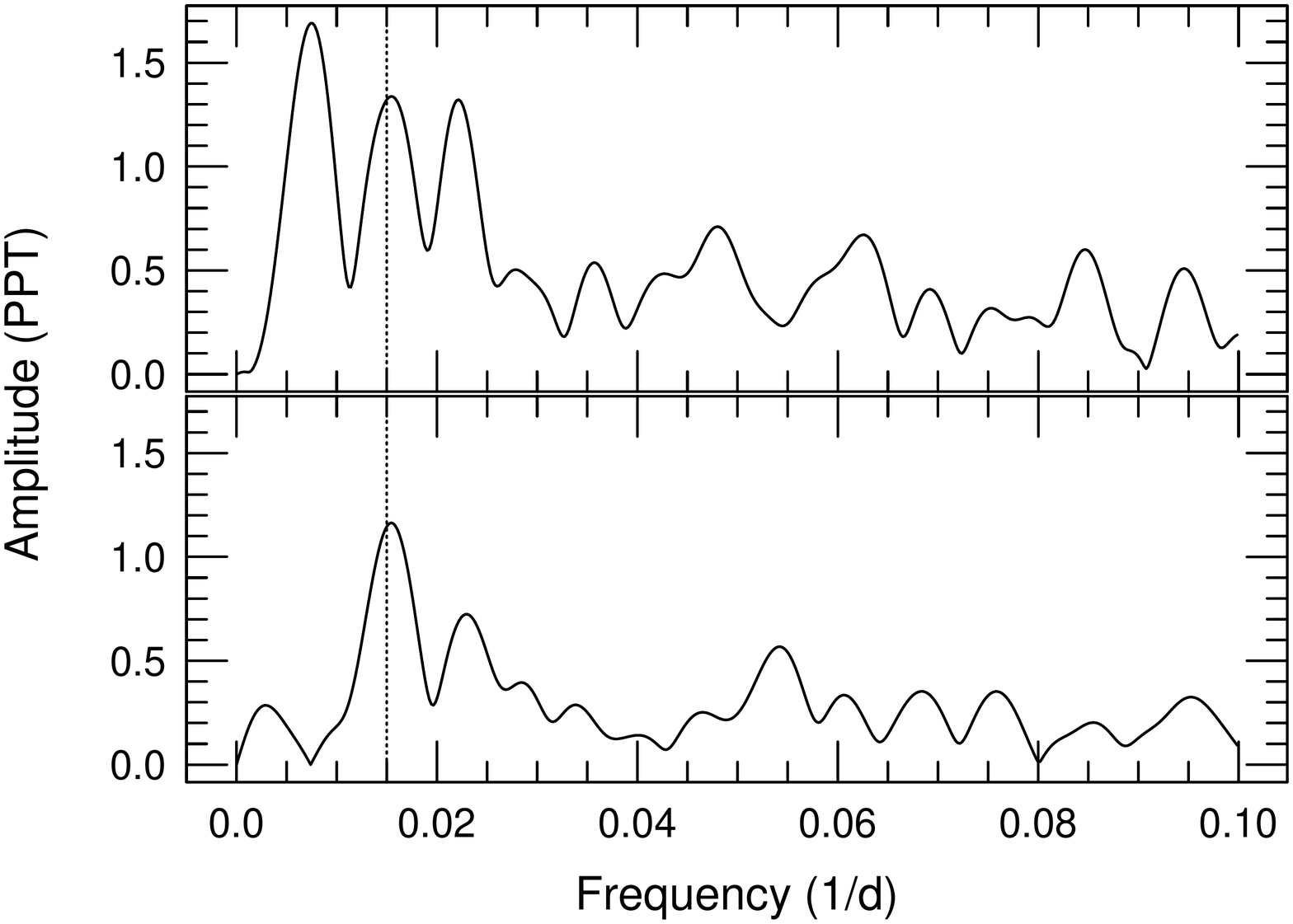}}
\caption{Pre-whitening procedure on the H$\alpha$ data taken between JD = 2455200 and 2455645.
The original amplitude spectrum before (top) and after (bottom)  removing the dominant frequency at 0.0078 d$^{-1}$.
The vertical dashed line marks the orbital frequency of GL~581d ($\nu$ = 0.015 d$^{-1}$, $P$ = 66.6 d).
}
\label{prewhite}
\end{figure}

The orbital frequency can also be easily detected  in a subset of the H$\alpha$ data taken 
JD = 2455200 and 2455645.  Because of the shorter data set, fewer Fourier components
were found, namely one at $\nu$ = 0.0079 d$^{-1}$ ($P$ = 126.9 d) and another
at  the orbital frequency of GL~581d, $\nu$ = 0.0150 d$^{-1}$ ($P$ = 66.4 d). Figure~\ref{prewhite} shows
the pre-whitening procedure on the subset $I_{H\alpha}$  data.
The removal of the dominant peak at $\nu$ = 0.0077 d$^{-1}$ ($P$ = 129 d)
results a strong peak at $\nu$ = 0.0152 d$^{-1}$, or $P$ = 65.8 d, near the orbital period of
GL~581d.

\begin{figure}[h]
\resizebox{\hsize}{!}{\includegraphics{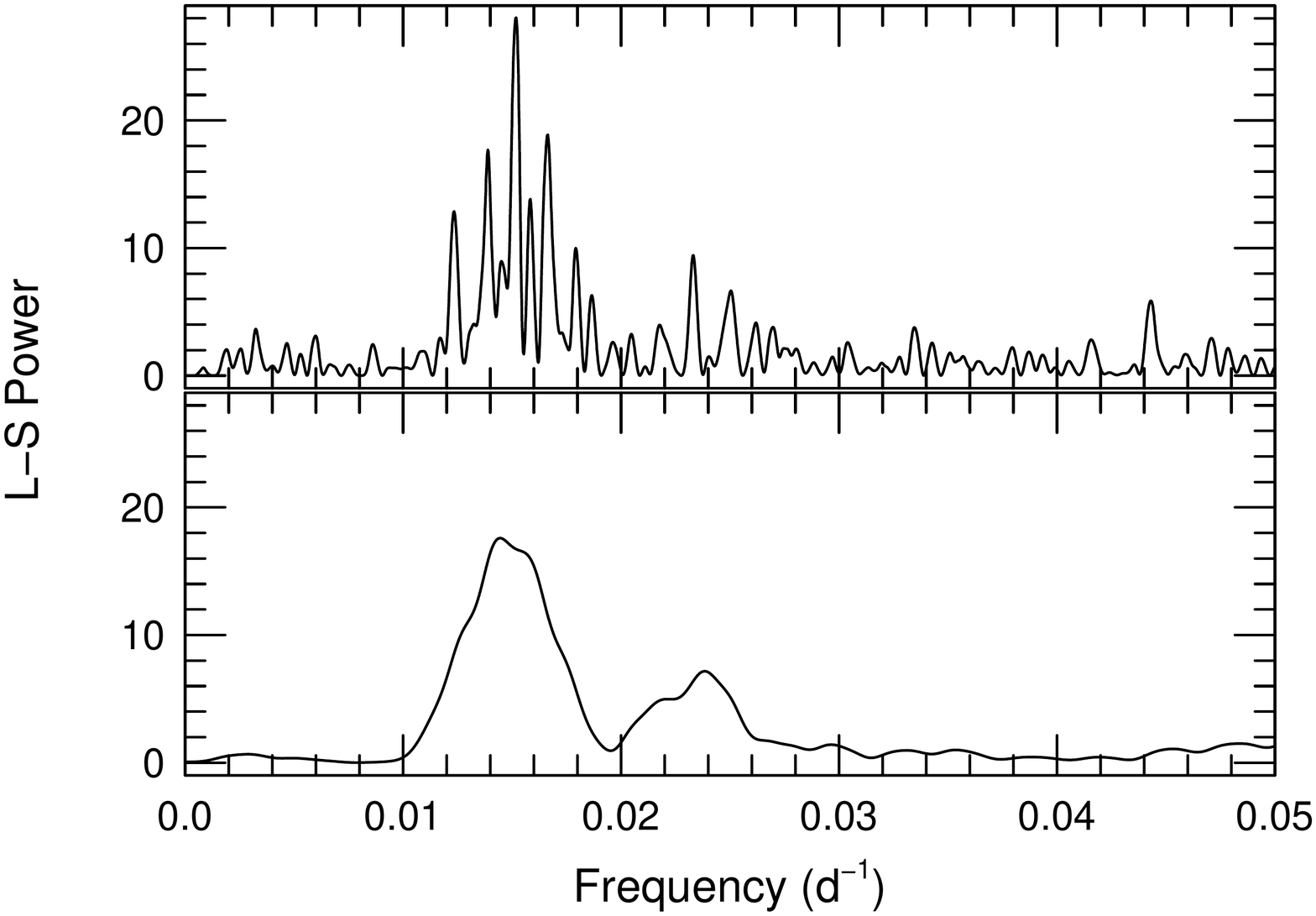}}
\caption{(top) The L-S periodogram of all  $I_{H\alpha}$  measurements after removing the contribution of all  frequencies
($f_1$, $f_3$ - $f_5$ in Table 1)  except that associated with GL~581d ($f_2$). (bottom)
The L-S periodogram of the  $I_{H\alpha}$  measurements taken between JD = 2455200 and 2455645
after removing the contribution of $f_6$ in Table 1. 
}
\label{scargle}
\end{figure}

\subsection{Statistical Significance of the Variations}

The statistical significance of a periodic signal can be assessed using the 
 Lomb-Scargle (L-S) periodogram (Lomb 1976; Scargle 1983). The top panel in Figure~\ref{scargle} shows the
L-S periodogram of the 
residual $I_{H\alpha}$ measurements after removing all frequencies except that associated with
GL~581d (i.e. $f_1$, $f_2$, $f_4$, and $f_5$ were removed). (Henceforth, we shall refer to ``residual'' $I_{H\alpha}$ measurements as those values where the contribution of all frequencies except $f_3$ have been removed from the data.)

One can estimate the false alarm probability (FAP) from the L-S power, $z$,   and the expression
FAP = 1 $-$ (1 $-$ $e^{-z}$)$^N$ $\approx$ $Ne^{-z}$, where $N$ is the number of independent frequencies (Scargle 1983). For the entire
data set this results in FAP $\approx$ 10$^{-13}$, For the subset data this is a FAP $\approx$ 10$^{-6}$. The latter value was confirmed
using a bootstrap procedure (Murdoch et al.  1993; K{\"u}rster et al. 1997) and 2$\times$10$^5$ random shuffles of the data. In no instance was
the L-S power of the random data greater than the actual value (FAP $<$ 5 $\times$ 10$^6$).

The L-S periodogram of the residual $I_{H\alpha}$ data from the subset data (JD = 2455200 -- 2455645)
is shown in the lower panel of Fig.~\ref{scargle}.  Note the dramatic
increase in the L-S power when using the full data set. This 
indicates a long-lived and reasonably coherent signal.
However, one should be wary of determining the FAP of a time series that has been modified. In this case the contributions
of four frequencies were removed from the data before estimating the FAP. If one removes the dominant peaks in the periodogram
the remaining ones will always look more significant.  Therefore we took additional approaches to estimate the FAP.

The L-S periodogram of the original time series before pre-whitening shows L-S power of $z$ $\approx$ 11.9 at the orbital frequency of
GL~581d ($\nu$ = 0.015 d$^{-1}$).  The above expression gives us the FAP for random noise producing more power than the actual data
over a broad frequency range. However, we are interested in assessing the FAP at a {\it known} frequency in the data, i.e. the
orbital frequency of GL~581d. In this case the FAP is given by FAP = $e^{-z}$ (Scargle 1983), or the previous expression with only
one independent frequency ($N$ = 1). This results in FAP $\approx$ 10$^{-5}$.

The FAP was assessed using a revised version of the bootstrap method. The five frequency components
in Table 1 were removed from the full $I_{H\alpha}$ index data. The resulting residuals represent our
``noise'' model for the data. The values from this noise model were then randomly shuffled keeping the
times fixed. The sine functions from the frequency analysis were then added back into into the noise data, except
for the contribution from GL~581d, $f_3$ (i.e. only
$f_1$, $f_2$, $f_4$, and $f_5$ were added back in). A L-S periodogram was calculated
and the maximum power in the frequency range $\nu$ = 0.01 -- 0.02 d$^{-1}$ found. We chose this frequency
range as we are interested in the probability noise would produce significant power at $\nu$ = 0.015 d$^{-1}$. With
2$\times$10$^5$ random shuffles of the data there was no instance when the random L-S periodogram produced power
higher than the original data. 

We also performed the same procedure on the subset of the $I_{H\alpha}$ index data. In this case there are fewer
Fourier components in the time series, namely the rotational frequency at $\nu$ = 0.0079 d$^{-1}$ 
and the orbital frequency of GL~581d, $\nu$ = 0.0150 d$^{-1}$. Both of these components were subtracted from the
data to produce the noise model for the bootstrap procedure. As before, the orbital frequency of GL~581d ($\nu$ = 0.0150 d$^{-1}$)
 was not added back
into the noise model prior to computing the L-S periodogram. Again, after 2 $\times$ 10$^{5}$ shuffles there was
no instance of the random periodogram having higher power than the actual data. A bootstrap analysis of the full and
subset data indicates that the FAP for the $I_{H\alpha}$  variations  at the orbital frequency of GL~581d is $<$
2 $\times$ 10$^{-6}$

The FAP can also be estimated directly from the Fourier amplitude spectrum and the height of a peak above the background
noise. Kuschnig et al. (1997) using Monte Carlo simulations established a relationship between the peak height above background
and the FAP (see Fig. 4 in their paper). 
In the case of the $I_{H\alpha}$ amplitude spectrum the peak at $\nu$ = 0.015 d$^{-1}$ is 4.3 times above the noise
level. This results in FAP $\approx$ 10$^{-4}$. Both the Fourier amplitude spectrum and the L-S periodogram analysis both indicate that
the $I_{H\alpha}$ variations at the orbital frequency of GL~581d are highly significant. 

\begin{figure}
\resizebox{\hsize}{!}{\includegraphics{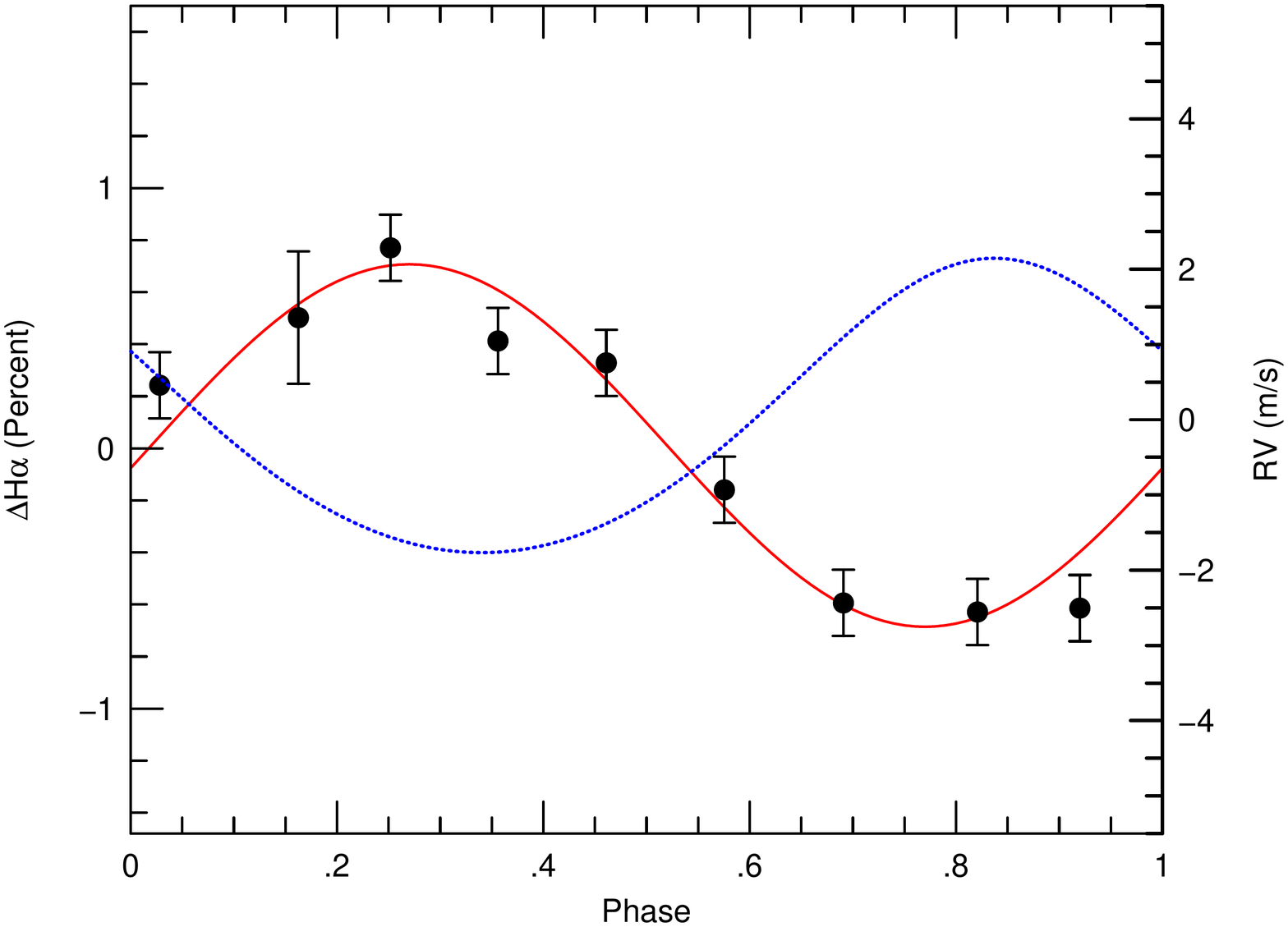}}
\caption{The phase-binned averages ($\Delta\phi$ $\approx$ 0.1) residual Ha variations phased to the 125-d period for ``GL~581d''. The solid curve 
represents a sine fit. The dashed curve represents the RV orbital solution for GL~581d.}
\label{habin}
\end{figure}

\section{$I_{H\alpha}$  versus RV variations}

Figure~\ref{habin} shows the residual $I_{H\alpha}$ variations phased to the orbital period of GL~581d determined by Hatzes (2014).
For clarity the data has been phase-binned on intervals $\Delta\phi$  $\approx$ 0.1. {The error bars
represent the standard deviation divided by the square root of the number of points in each bin.} The solid curve represents
a sine fit to the data. Also shown as a dashed line is the RV orbital solution for GL~581d. The $I_{H\alpha}$ -- RV variations
for GL~581d are anti-correlated thus supporting the conclusions of Robertson et al. (2014). The variations
can  be fit by a pure sine function. If the RV variations attributed to GL~581d are actually due to 
activity, then these should mimic a circular Keplerian orbit. There appears to be a slight phase shift of $\approx$ 0.1
between the RV and $I_{H\alpha}$ data but this is not deemed significant and may be an artifact of the binning process. 

\begin{figure}
\resizebox{\hsize}{!}{\includegraphics{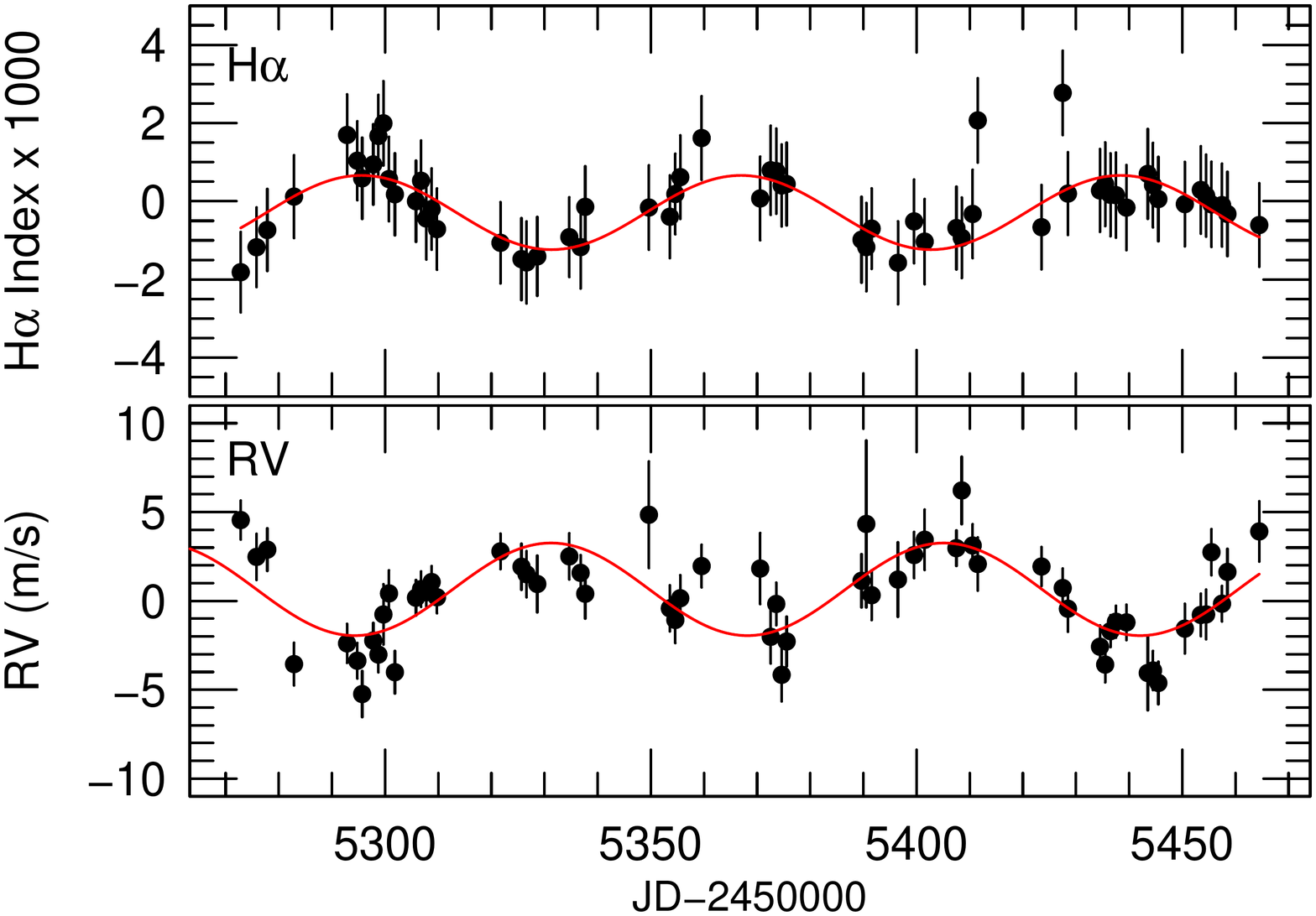}}
\caption{(Top) The variations in $I_{H\alpha}$ over the time span JD-2450000 = 5272 -- 5464. The curve represents a sine
fit with the orbital period of GL~581d (Bottom) The RV variations of the purported planed ``d'' (all other planet signals removed)
over the same time span. The curve represents the orbital solution. }
\label{harv}
\end{figure}

Over the interval 
 JD = 2455200 and 2455645 the sampling of the data (both RV and $I_{H\alpha}$)  was excellent and almost
 three cycles of variations were covered. The residual $I_{H\alpha}$  data during this time
 interval  are shown in Figure~\ref{harv} along with the RV variations 
 due to GL~581d (i.e. the contribution of all other planets and activity have been removed).  
 Note that in this case the RV and $I_{H\alpha}$ variations are exactly
 180$^\circ$ out-of-phase with each other. 
 
A sine fit to the RV and $I_{H\alpha}$ variations over this interval  and allowing the period to be a free
parameter resulted in consistent values for the period for the two quantities, namely
71.99 $\pm$ 1.51 d and 70.64 $\pm$ 1.64 d for the H$\alpha$ and
RV, respectively. These periods are slightly longer by about 1.4$\sigma$
compared to  the orbital period of 66.64 $\pm$ 0.08 d derived using the full data set. This may be an  indication
of possible period variations due to differential rotation which may also be consistent with activity-related
variations. Unfortunately, we cannot be sure given the short time span of the data. 

We also examined the ratio of  the amplitudes of the $I_{H\alpha}$ and RV variations ($A_{H\alpha}$/$A_{RV}$). This ratio
should be constant if the two variations stemmed from the same origin. However, we might see temporal variations
if the RV amplitude is constant and due to a planet, while the activity signal from $I_{H\alpha}$ varied due to the
evolution of the stellar active regions.

\begin{figure}
\resizebox{\hsize}{!}{\includegraphics{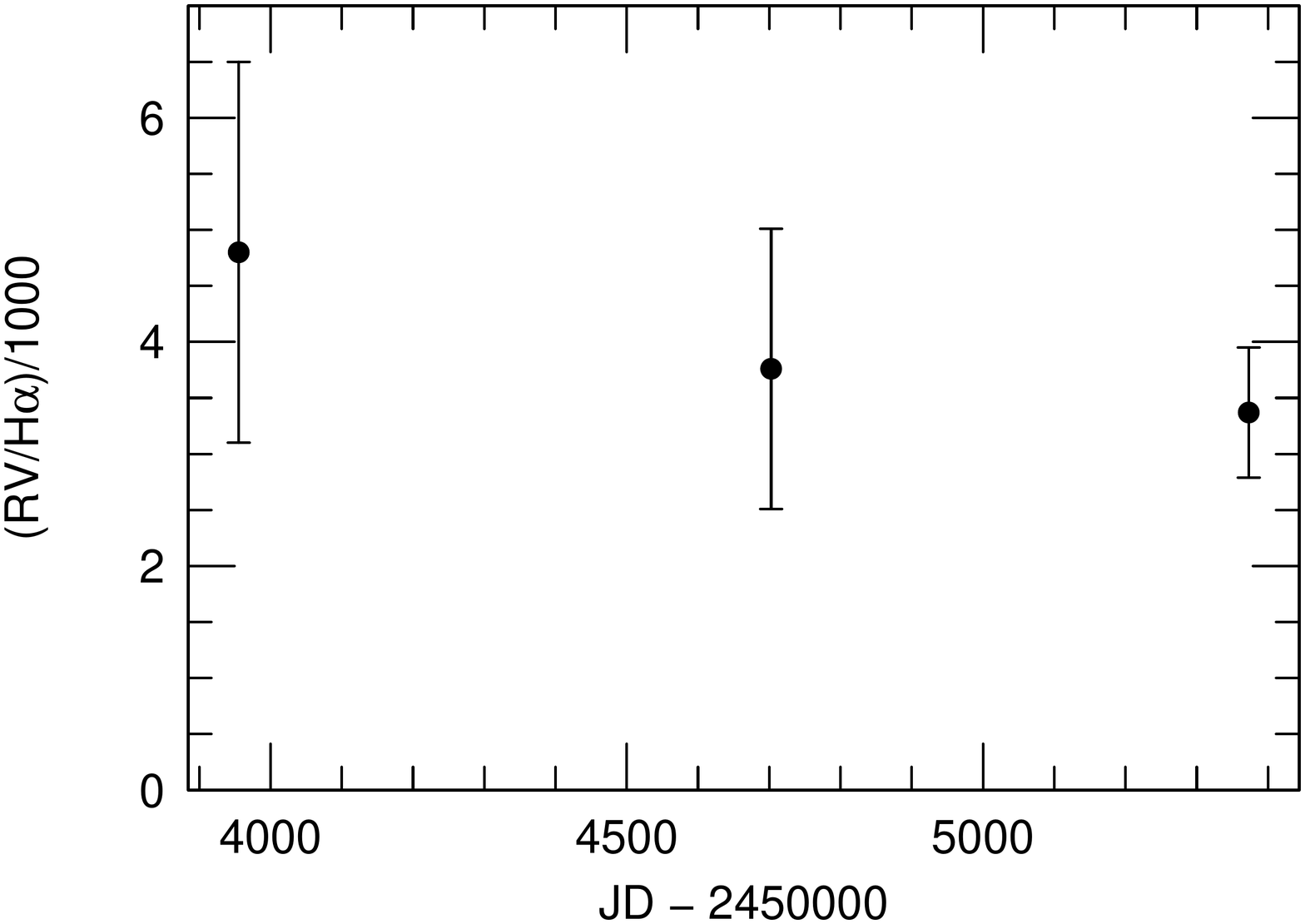}}
\caption{The ratio of the RV to $I_{H\alpha}$ amplitude ($I_{H\alpha}$ amplitude is in PPT) for three different epochs. }
\label{ratio}
\end{figure}

Figure~\ref{ratio} shows the ratio $A_{H\alpha}$/$A_{RV}$ for three different epochs. The $I_{H\alpha}$ amplitude was calculated in
parts per thousand (PPT). Although there appears to be a slight
decrease in the ratio with time, this is not significant. To within the errors the ratio $A_{H\alpha}$/$A_{RV}$ is
constant with time.  {However, given that the active regions may evolve at different time scales to those on sun-like stars
the constancy of $A_{H\alpha}$/$A_{RV}$ be less informative due to the restricted time base of the measurements.}

\begin{table}[h]
\begin{center}
\begin{tabular}{cccc}
Label   & Frequency &  Period & Amplitude  \\
             & (d$^{-1}$)   &  (d)   &  (PPT)  \\
\hline
$f_1$ & 0.00719 $\pm$ 0.000028 & 139.16 $\pm$ 0.42 & 0.883 $\pm$  0.079 \\
$f_2$ & 0.00797 $\pm$ 0.000033 & 125.39 $\pm$ 0.45 & 0.642 $\pm$  0.071  \\
$f_3$ & 0.01517 $\pm$ 0.000035 & 65.91  $\pm$ 0.44 & 0.618 $\pm$  0.063  \\
$f_4$ & 0.00037 $\pm$ 0.000058 & 2674 $\pm$ 414    & 0.568 $\pm$  0.100  \\
$f_5$ & 0.02282 $\pm$ 0.000051 & 43.83 $\pm$ 0.08  & 0.376 $\pm$  0.006  \\
\end{tabular}
\caption{Frequencies found in the $I_{H\alpha}$ data}
\end{center}
\label{parameters}
\end{table}

\section{Discussion}

Using a pre-whitening procedure we were able to isolate the variations of $I_{H\alpha}$ in GL~581 at a
frequency of $\nu$ = 0.015 d$^{-1}$, $P$ = 66.6 d which is coincident the RV variations attributed to the planet 
GL~581d. These variations show sinusoidal variations that are 180$^\circ$ out-of-phase with the ``orbital'' RV variations due
to GL~581d. This confirms the anti-correlation between the RV variations of GL~581d and $I_{H\alpha}$  found by
Robertson et al. (2014).  The RV variations attributed to GL~581d are most likely due to stellar activity. 
{GL~581d represents another case where activity-related RV variations can mimic a Keplerian orbit
(see Queloz et al. 2001; Bonfils et al. 2007). }

The $I_{H\alpha}$ variations appear to be long-lived since they are visible in data spanning almost 7 years. This is certainly a cautionary
tale for RV programs searching for planets in the habitable zone around M-dwarf stars.

Of course, the presence of the period of GL~581d in $I_{H\alpha}$ is no proof that the planet does not exist.
There is no physical reason why a planet cannot have the same
orbital period as the stellar rotation, or one of its harmonics. There are two arguments against keeping
GL~581d as a confirmed exoplanet. First, when one finds significant periodic variations in an activity indicator
 (photometry, $I_{H\alpha}$, Ca II, bisectors, etc.)
with the same period as the RV variations then this is generally accepted as a  ``non-confirmation'' of the planet
candidate. There is no reason to change this criterion simply to ``save'' a habitable planet. It is better to exclude a questionable
exoplanet in the census rather than keeping it when it actually is not there.

Second, Robertson et al. (2014) demonstrated that after correcting the RVs due to their correlation with $I_{H\alpha}$ this boosted
the significance of the other RV signals that are certainly due to planetary companions. This behavior is also  consistent
with activity-related RV variations for the 66.7 d signal.

We examined the  the ratio of the  RV to $I_{H\alpha}$ amplitudes. If the RV amplitude was
directly tied to the activity then this ratio should remain roughly
constant. On the other hand, if the 66.7 d RV variations were due to a planet then the amplitude
of these would remain
constant, whereas an activity cycle may produce amplitude variations in the $I_{H\alpha}$  index. 
This should result in a variation in the  RV to $I_{H\alpha}$ amplitude and the planet candidacy 
of GL~581d could remain.
Although our measurements of the RV to $I_{H\alpha}$ amplitude is consistent with no
variations, the measurement error is too large to exclude with certainty any temporal variations. 
Unfortunately, to detect any temporal variations in either the $I_{H\alpha}$ or RV amplitude would require an inordinate amount of additional measurements. At the present time there is no compelling reason to think that  GL~581 is more than a  3-planet system.


\clearpage

\end{document}